\documentclass[
 superscriptaddress,
 nofootinbib,
 nobibnotes,
 amsmath,amssymb,
 aps,
 prd,
 twocolumn, 
]{revtex4-2}

\usepackage{graphicx}
\usepackage{dcolumn}
\usepackage{bm}
\usepackage{hyperref}
\usepackage{appendix}
\usepackage{mathtools}
\usepackage{multirow}
\usepackage{amsmath,amsfonts,bm}
\usepackage{wasysym}
\usepackage{xcolor}
\usepackage{xspace}
\usepackage{lineno}

\begin{document}

\title{TLS loss in superconducting resonators on epitaxial hBN thin films}

\author{J.~Christiansen-Salameh}\thanks{Now at Conductor Quantum, San Francisco, CA 94114, USA}
\affiliation{Sibley School of Mechanical and Aerospace Engineering, Cornell University, Ithaca, NY 14853, USA}
\affiliation{Quantum Division, Fermi National Accelerator Laboratory, Batavia, IL 60510, USA}

\author{D.J.~Temples}
\affiliation{Quantum Division, Fermi National Accelerator Laboratory, Batavia, IL 60510, USA}

\author{R.~Linehan}
\affiliation{Quantum Division, Fermi National Accelerator Laboratory, Batavia, IL 60510, USA}
\affiliation{Department of Physics and Astronomy, Northwestern University, Evanston, IL 60208, USA}

\author{S.~Sussman}
\affiliation{Quantum Division, Fermi National Accelerator Laboratory, Batavia, IL 60510, USA}

\author{Aloyse Maille}
\affiliation{Sibley School of Mechanical and Aerospace Engineering, Cornell University, Ithaca, NY 14853, USA}

\author{F.~Abouzahr}
\affiliation{Department of Physics and Astronomy, Northwestern University, Evanston, IL 60208, USA}


\author{L.~Alexis}
\affiliation{Sibley School of Mechanical and Aerospace Engineering, Cornell University, Ithaca, NY 14853, USA}

\author{K.~Anyang}
\affiliation{Department of Physics, Illinois Institute of Technology, Chicago, IL 60616, USA}

\author{D.~Bowring}
\affiliation{Quantum Division, Fermi National Accelerator Laboratory, Batavia, IL 60510, USA}

\author{G.~Bratrud}
\affiliation{Department of Physics and Astronomy, Northwestern University, Evanston, IL 60208, USA}
\affiliation{Quantum Division, Fermi National Accelerator Laboratory, Batavia, IL 60510, USA}


\author{E.~Celi}
\affiliation{Department of Physics and Astronomy, Northwestern University, Evanston, IL 60208, USA}

\author{A.H.~Colón~Cesaní}
\affiliation{Department of Physics and Astronomy, Northwestern University, Evanston, IL 60208, USA}



\author{R.~Khatiwada}
\affiliation{Department of Physics, Illinois Institute of Technology, Chicago, IL 60616, USA}
\affiliation{Quantum Division, Fermi National Accelerator Laboratory, Batavia, IL 60510, USA}

\author{D.~Molenaar}
\affiliation{Department of Physics, Illinois Institute of Technology, Chicago, IL 60616, USA}
\affiliation{Quantum Division, Fermi National Accelerator Laboratory, Batavia, IL 60510, USA}


\author{O.~Peek}
\affiliation{Department of Materials Science and Engineering, Cornell University, Ithaca, NY 14853, USA}

\author{A.~Rodriguez}
\affiliation{Department of Physics and Astronomy, Northwestern University, Evanston, IL 60208, USA}

\author{O.~Seidel}
\affiliation{Department of Physics, University of Texas at Arlington, Arlington, TX 76019, USA}
\affiliation{Quantum Division, Fermi National Accelerator Laboratory, Batavia, IL 60510, USA}

\author{D.~Schlom}
\affiliation{Department of Materials Science and Engineering, Cornell University, Ithaca, NY 14853, USA}
\affiliation{Platform for the Accelerated Realization, Analysis, and Discovery of Interface Materials (PARADIM), Cornell University, Ithaca, NY 14853, USA}

\author{D.~Sotir}
\affiliation{Department of Materials Science and Engineering, Cornell University, Ithaca, NY 14853, USA}

\author{S.~Uemura}
\affiliation{Quantum Division, Fermi National Accelerator Laboratory, Batavia, IL 60510, USA}

\author{Y.~Wu}
\affiliation{Sibley School of Mechanical and Aerospace Engineering, Cornell University, Ithaca, NY 14853, USA}

\author{E.~Figueroa-Feliciano}
\affiliation{Department of Physics and Astronomy, Northwestern University, Evanston, IL 60208, USA}
\affiliation{Quantum Division, Fermi National Accelerator Laboratory, Batavia, IL 60510, USA}

\author{D.~Baxter}\thanks{dbaxter9@fnal.gov}
\affiliation{Quantum Division, Fermi National Accelerator Laboratory, Batavia, IL 60510, USA}
\affiliation{Department of Physics and Astronomy, Northwestern University, Evanston, IL 60208, USA}

\author{Z.~Tian}\thanks{zt223@cornell.edu}
\affiliation{Sibley School of Mechanical and Aerospace Engineering, Cornell University, Ithaca, NY 14853, USA}




\date{Received: date / Accepted: date}

\begin{abstract}
The integration of low-loss dielectrics into superconducting circuits is of importance to the functionality of quantum computers and sensors. In this work, we fabricate and measure niobium resonators atop few-layer, epitaxial hexagonal boron nitride (hBN) grown by Molecular Beam Epitaxy on a sapphire substrate and compare against a control counterpart without hBN. We conduct radiofrequency measurements which probe the responses to power, temperature, and temporal fluctuations of two-level systems. The resonators integrating hBN maintain low dielectric loss with a low-power internal loss tangent on the order of $10^{-6}$, and demonstrate hBN can be added to the conventional superconducting qubit stack with minimal degradation of the quality factor. This research represents a first step towards a materials-driven strategy to eliminate phonon-mediated correlated errors in solid-state quantum systems, which leverages the thermal properties of 2D materials to prevent athermal phonons excited in the device substrate from transmitting across the superconductor-substrate interface. The same technique applied to quantum sensors would maximize phonon collection by blocking phonons from entering passive metal films.
\end{abstract}

\maketitle

\section{Introduction}
The functionality of superconducting circuits can be compromised by excess quasiparticle (QP) bursts caused by non-equilibrium phonon excitations~\cite{2004PhRvL..92f6802A,Wang:2014hnf,DeVisser2014,Serniak:2018itt,Vepslinen2020}. Gamma-rays emitted by radioactive impurities and cosmic-ray muons deposit energy primarily in the device substrate due to its large interaction volume relative to the superconducting thin film. This radiation triggers a cascade that efficiently converts the initial energy into phonons~\cite{Martinis2021} with energies orders of magnitude larger than the millikelvin thermal bath. Athermal phonons travel ballistically throughout the substrate, and some fraction transmit across the substrate-superconductor interface with sufficient energy ($>$2$\Delta$, where $\Delta$ is the superconducting gap) to break Cooper pairs, producing excess QPs~\cite{Kaplan} that cause decoherence. Similar bursts may occur from intrinsic relaxation of mechanical stress from device mounting and lattice mismatch~\cite{Anthony-Petersen2022, Baxter:2025odk, Yelton:2025wsy}.

In quantum computing applications, the phenomenon of QP-induced decoherence is dubbed ``quasiparticle poisoning.” The dissipative tunneling of QPs across the Josephson junction of a superconducting qubit causes energy relaxation and dephasing at a rate proportional to the QP density~\cite{2011arXiv1101.0862L}. Qubit errors mediated by phonons are especially detrimental because they can be correlated, impacting multiple qubits over large time (ms) and length scales (cm) due to the long lifetimes of acoustic phonons in the device substrate~\cite{Martinis2021}. Existing efforts to mitigate phonon-mediated quasiparticle poisoning include radiation shielding~\cite{Cardani2021,Bratrud2025}, gap engineering~\cite{Sun:2011kyo,Diamond:2022scj,Mcewen2024,Harrington:2024iqm}, and phonon traps~\cite{Henriques2019,Karatsu:2019qpl,Iaia2022}. 

In quantum sensing applications, the transduction of energy initially deposited in a substrate into the superconductor QP population is a mechanism for detection. For example, Kinetic Inductance Phonon-Mediated (KIPM) detectors built from superconducting resonators leverage the QP-density-dependent change in kinetic inductance and corresponding shift in the resonant frequency and quality factor to identify pulses associated with particle impacts~\cite{Temples2024,Golwala2008, Golwala2022, Moore2012, Cardani2018, Wen2022}. A KIPM detector’s sensitivity is limited in part by both RF noise and the loss of phonons to inactive superconductor (e.g. a ground plane) or the surrounding device packaging, limiting the phonon collection efficiency to 1--25\%~\cite{11478312, Cruciani:2022mbb}. 

\begin{figure}[t]
        \includegraphics[width=0.49\textwidth]{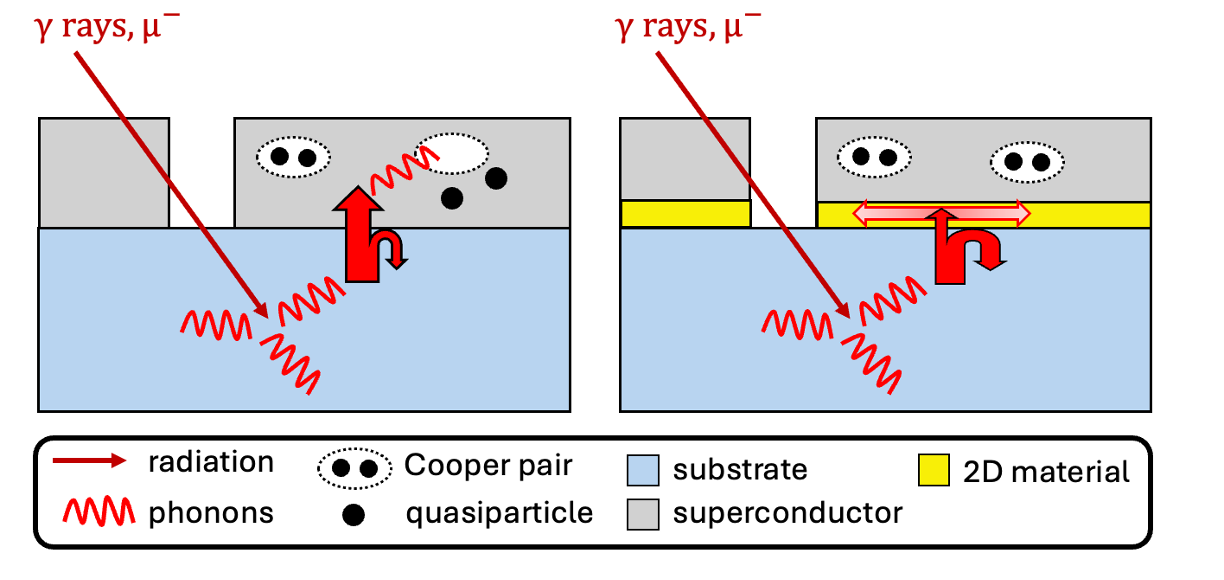}
    \caption{Diagram of a phonon-mediated quasiparticle burst in a superconducting resonator. We hypothesize that the insertion of a 2D material at the superconductor-substrate interface, as indicated in yellow in the right diagram, will reflect and diffuse athermal phonons originating from the substrate to prevent Cooper pair-breaking and protect the device from correlated decoherence.}
    \label{fig:hbn_diagram}
\end{figure}

We posit that a 2D material inserted at the superconductor-substrate interface could be patterned strategically to suppress interfacial phonon transmission for either efficient mitigation of phonon-mediated QP poisoning in computing applications or an enhancement of the phonon collection efficiency in sensing applications by blocking energy loss. 2D materials have highly-anisotropic thermal transport properties~\cite{Jaffe2023, Jiang2018, Kim2021, Luo2015}. In the cross-plane, weak Van der Waals forces yield a small phonon transmission coefficient~\cite{Foss2021, Lu2025,Suryavanshi2019}, meaning the majority of phonons will be reflected at the interface. Strong covalent bonding gives a high thermal conductivity in-plane, which will diffuse laterally the fraction of phonons that are transmitted~\cite{Cai2019, Balandin2008}. Figure~\ref{fig:hbn_diagram} illustrates the concept of using 2D materials to protect regions of a superconducting device from quasiparticle bursts. We identify hexagonal boron nitride (hBN) as an ideal candidate due to its low dielectric loss tangent in exfoliated single-crystals~\cite{Wang2022}.

When introducing a new material into the materials stack of a superconducting device, a critical challenge is preventing the introduction of dielectric loss, which is large at material interfaces. 
Material defects such as amorphous grain boundaries, impurity atoms, vacancies, etc. are hosts for two-level systems (TLS)~\cite{Mller2019}. TLS create loss by resonantly absorbing the energy stored in the superconducting circuit~\cite{Mller2019}. As a result, TLS loss has been found to limit the quality factors of superconducting resonators and qubits alike~\cite{Wang2020, Martinis2005}, even without incorporating an additional 2D material layer. Additionally, TLS manifest in the noise spectra and therefore limit the sensitivity of detectors, such as KIPM detectors~\cite{Temples2024}. Therefore, the method of preparing the hBN layer must be selected to promote cleanliness and excellent crystallinity to minimize additional TLS. 

Bulk, single-crystal dielectric substrates such as sapphire or high-resistivity silicon have low dielectric loss tangents ($\tan\delta$) of order $10^{-8}$ and $10^{-6}$, respectively~\cite{Read2023, Checchin2022}. Flakes of hBN exfoliated from bulk crystals have $\tan\delta$ of $5 \times 10^{-6}$ in the single-photon regime~\cite{Wang2022}. However, exfoliated flakes suffer limited area, low yield, and poor reproducibility. Alternatively, thin film deposition of hBN offers coverage of the full area of the chip, scalability, and compatibility with standard device fabrication workflows. Deposited thin film dielectrics cannot achieve the same crystallinity and purity of bulk crystals, and generally have $\tan\delta$ in the range $10^{-2}$ to $10^{-5}$~\cite{Wang2020}. For example, order 100\,nm thickness of disordered BN sputtered on superconducting resonators has zero-temperature TLS loss of $6\times 10^{-5}$~\cite{Wisbey2019}. We note a gap in the literature between these extremes of exfoliated single-crystal hBN and sputtered, amorphous BN. In this work, we integrate ordered hBN thin films at the superconductor-substrate interface of radiofrequency (RF) resonators and evaluate the resulting losses to inform the feasibility of using hBN to suppress the phonon mediation of quasiparticle bursts in computing architectures or to block the loss of energy in passive films in quantum sensors. 

\section{Methods}\label{s:methods}
High-temperature molecular beam epitaxy (MBE) with laser substrate heating was used to produce wrinkle-free, epitaxial, few-layer hBN thin films rotationally aligned to c-plane sapphire substrates. MBE is an ultra-high vacuum technique with a slow and controlled growth rate, which promotes film purity and crystallinity. Furthermore, local substrate heating by the laser heater mitigates the thermal migration of contaminants from the assembly to substrate compared to conventional resistance heating. An open-source device layout featuring eight coplanar waveguide resonators coupled to a common feedline~\cite{Kopas2022} was selected to enable 1:1 comparison of material stacks and fabrication workflows, and the participation ratios were designed for sensitivity to dielectric loss arising from the metal-substrate interface. Nb resonators were fabricated both on the MBE-grown hBN thin films and on bare sapphire substrates, designated as the hBN and control chips, respectively. RF measurements to probe the TLS-driven responses to power, temperature, and stochastic fluctuations in time were conducted to thoroughly compare loss in the resonators on hBN to the control. 

\subsection{Molecular Beam Epitaxy}
Thin films of hBN were grown on sapphire by MBE at the Platform for the Accelerated Realization, Analysis, and Discovery of Interface Materials. Further details of the growth and characterization of epitaxial hBN thin films are reported in Appendix \ref{sec:app0} and Ref.~\cite{accompanyingwork}. For the purpose of the measurements outlined here, the \textit{in situ} Resonant High Energy Electron Diffraction pattern developed diffuse vertical streaks, evidencing epitaxial thin film growth and rotational alignment to the underlying sapphire substrate. The Raman spectra displayed the characteristic hBN peak near 1372\,cm$^{-1}$. Cross-sectional scanning transmission electron microscopy imaging revealed well-ordered layers of hBN with estimated 3\,nm total thickness. Finally, atomic force microscopy mapping showed the thin film surface is wrinkle-free with 0.3\,nm RMS surface roughness. 

\subsection{Resonator Fabrication and Packaging}
Nb thin films were deposited on the MBE-grown hBN thin films and blank c-plane sapphire substrates using an AJA sputtering system. An outgassing and chamber seasoning procedure was followed to limit impurities incurred by use of a shared system that also sputters magnetic materials. An RF clean step was omitted and the DC power was reduced to 50\,W to protect the hBN thin film from damage via physical bombardment (see Figure~\ref{fig:TEM} of Appendix \ref{sec:app1}). A Nb thickness of 61\,nm was measured with a P7 contact profilometer on a witness wafer, establishing a Nb sputtering rate of 0.29\,\AA/s. The hBN and control devices were sputtered in the same batch to mitigate variation in Nb film purity or thickness. The resonators were patterned into the Nb film by photolithography and chlorine plasma etching.  The etch rate of hBN under the conditions provided in Appendix \ref{sec:app1} was measured to be 3 times slower than the Nb etch rate. Given the few-layer thickness of the hBN, it is likely the hBN was entirely etched where exposed to the plasma and remained only beneath the regions of Nb protected by photoresist (\textit{i.e.} hBN was removed from the resonator and feedline gaps). 
The device packaging consisted of a printed circuit board and Cu housing. The packaging design has been used in Ref. \cite{Huang2025} and is similar to designs reported in Ref. \cite{Huang2021} (see Appendix \ref{sec:app2} and Figure~\ref{fig:device}).

\subsection{Experimental Setup}

Both devices were measured simultaneously in the Quantum Underground Instrumentation Experimental Testbed (QUIET) at Fermi National Accelerator Laboratory, a deep-underground facility with 100\,m of rock overburden that shields $>$99$\%$ of cosmic ray muons. QUIET is equipped with an Oxford Proteox-MX dry Dilution Refrigerator. The two packaged devices were installed on the same Cu bar thermally connected to the bottom of the mixing chamber plate and enclosed within a double-walled Amuneal A4K shield. Figure~\ref{fig:fridge_diagram} of Appendix \ref{sec:app3} shows the components of the RF input and output lines with an estimated 77\,dB of attenuation from the RF signal generator to the device. The various $S_{21}$ transmission measurements presented here were taken using a Keysight E5080B ENA Vector Network Analyzer (VNA), a Quantum Instrumentation Control Kit (QICK)~\cite{Stefanazzi2022} on a Xilinx RFSoC 216-board, or an Ettus Research x300 USRP streaming DAQ.

\section{Results and Discussion}

\begin{figure}[t]
        \includegraphics[trim={0 2pt 2pt 0}, clip, width=0.49\textwidth]{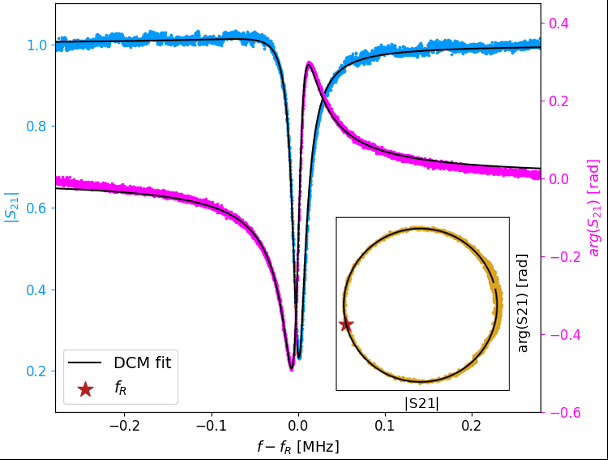}
    \caption{Representative example of a resonance at $f_R = 4.265 $\,GHz and the Diameter Correction Method (DCM) fit (black line) of the linear magnitude (orange data points) and phase (magenta data points) of the $S_{21}$ parameter. The inset shows the DCM fit in the complex plane with a red star designating $f_R$.}
    \label{fig:example_resonance}
\end{figure}

Resonances were first identified in a frequency sweep of the forward transmission coefficient ($S_{21}$) measured with the VNA. The diameter correction method (DCM) ~\cite{Khalil2012} was adapted from the \texttt{scresonators} ~\cite{scresonators} package and applied to fit $S_{21}$$(f)$ in the complex plane, from which the resonant frequency ($f_R$), loaded quality factor ($Q$), internal quality factor ($Q_i$), and coupling quality factor ($Q_c$) were extracted. An example of the spectroscopy data for one resonator is shown in Figure~\ref{fig:example_resonance}. Base temperature measurements of $f_R$ and $Q_c$ of all eight resonators of each chip are listed in Table~\ref{tab:placeholder} of the Appendix. The internal loss is defined to be $\delta_i = 1/Q_i$, and as such can be extracted directly from the DCM fit. 

\subsection{Power Sweep}
\begin{figure}[t]
        \includegraphics[width=0.49\textwidth]{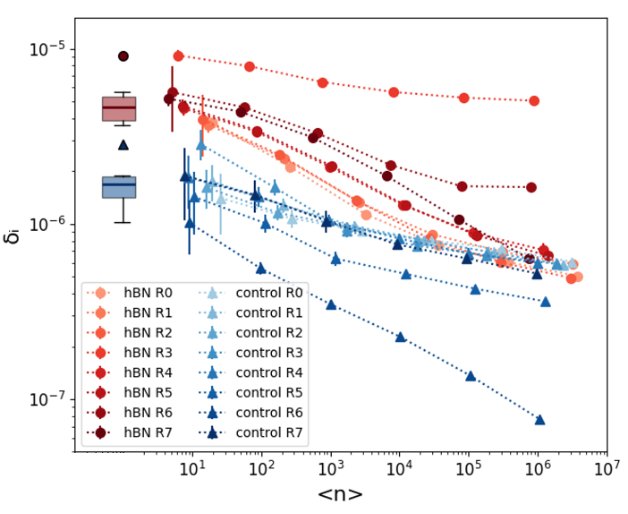}
    \caption{Power-dependence of the internal loss ($\delta_i$) of hBN (red circles) and control (blue triangles) resonators as a function of $\langle n\rangle$, the average photon number in the resonator. The error bars represent the 95\% confidence interval of the DCM fit. The box plots compile the low-power internal loss ($\delta_i^{\mathrm{LP}}$) across the set of resonators on the hBN (red) and control (blue) chips, and are constructed from only the lowest power measurement for each control or hBN resonator, which vary in $\langle n\rangle$ due to unique values of $Q$ for each resonator. }
    \label{fig:power_sweep}
\end{figure}

TLS facilitate loss by absorbing a microwave photon from the resonator and emitting a phonon to the bath. Strong microwave drive excites transitions to the higher energy level(s), saturating TLS and reducing the net absorption of photons from the resonator. Therefore TLS loss decreases with increasing drive power, with the underlying physics well-understood (see Ref.~\cite{Pappas2011,Gao2008} for more details). The internal component of the loss arises from the dielectric properties of the materials stack including effects of TLS, and is an indicator of the merit of the materials selection and fabrication processes. Specifically, the low-power internal loss ($\delta_{i}^\mathrm{LP}$) in the single-photon regime represents the operating conditions of resonators for readout and control of superconducting qubits, and is on the end of the power spectrum most susceptible to TLS loss. Resonator spectroscopy data was taken with the VNA at various drive powers. Figure~\ref{fig:power_sweep} shows the power-dependent $\delta$$_i$ of the hBN and control resonators as a function of $\langle n\rangle$, the average number of photons in the resonator.  $\langle n \rangle$ was estimated from the the input RF drive power, the total attenuation on the input line, and the measured parameters of each resonator \cite{Wang2020}. 

We take our $\delta_i^{\mathrm{LP}}$ as on the order of single-photon power. Low signal-to-noise ratio prevented reliable fitting of resonances at $\langle n\rangle \le$ 1, and we did not observe the flattening out of $\delta_i$ at low $\langle n \rangle$ (low readout power) indicative of approaching (from high $\langle n \rangle$ to low $\langle n \rangle$) the characteristic photon number of the onset of TLS saturation~\cite{Pappas2011,Gao2008}. Nonetheless, the reported $\delta_i^{\mathrm{LP}}$ is at sufficiently low power to represent the operating conditions of superconducting qubits. The hBN resonators have elevated $\delta_i^{\mathrm{LP}}$ with a median of $4.50 \times 10^{-6}$ and a standard deviation ($\sigma$) of $1.7 \times 10^{-6}$, compared to the control resonators with a median of $1.65 \times 10^{-6}$ and $\sigma$ of $0.50 \times 10^{-6}$, though remain on the same order of magnitude. For $\langle n \rangle > 10^{4}$, $\delta_i$ converges for the majority of resonators to a similar trend, regardless of the presence of hBN, consistent with TLS saturation.
Across each of the control and hBN chips, $\delta_i$ is not found to correlate with $f_R$, and may instead stem from uncontrolled variations in the fabrication processes across the area of the chip. A possible explanation for the larger spread in $\delta_i^{\mathrm{LP}}$ on the hBN chip is the distribution of crystalline defects in the hBN thin film; for example, a region with stacking faults aligning underneath a resonator trace could introduce more loss. The control resonators have $\delta$$_i^{\mathrm{LP}}$ on par with state-of-the-art Nb CPW resonators, with the caveat of unaccounted differences in geometry, substrate materials, fabrication processes, magnetic shielding, IR filtering, thermalization, etc. As one example, high-Q Nb-on-Si resonators of the same device layout achieved a median $\delta_i^{\mathrm{LP}}$ of $1.2\times 10^{-6}$~\cite{Olszewski:2025rxc}. Extending to resonators with geometries other than Ref.~\cite{Kopas2022}, the median $\delta_i^{\mathrm{LP}}$ is in the range ($0.1$--$3) \times 10^{-6}$~\cite{Wang2020, Olszewski:2025rxc, Drimmer2024,Oh2024,Torres-Castanedo2024,Zheng2022,Alto2022,Alghadeer2022,Kowsari2021,Manzo-Perez2025,Wang2024}. Our hBN resonator $\delta_i^{\mathrm{LP}}$ remains on the same order of magnitude of this range. 

\begin{figure*}[t!]
        \includegraphics[width=0.99\textwidth]{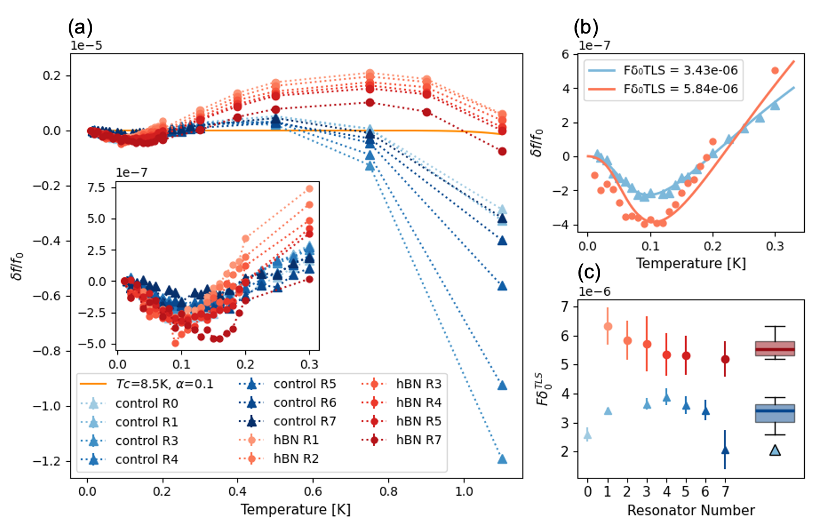}
    \caption{ Temperature-dependence of fractional frequency shift of hBN (red circles) and control (blue triangles) resonators. (a) $\delta f(T)/f_0$ across the measured temperature span up to 1.1 K. The error bars show the 95\% confidence interval of the DCM fit and are obscured by the marker size. The orange line demonstrates negligible contribution to $\delta f/f_0$ from thermal quasiparticles as predicted by Mattis-Bardeen theory. The inset shows the TLS-dominated temperature regime below 300 mK used for fitting. (b) Representative fits to Equation~\eqref{temp_eq} where $T_c$ is set to 8.9 K and $\alpha$ is set to 0.1. The selected resonators shown are control R1 (blue triangles) and hBN R2 (red circles). (c) Extracted $F\delta_0^{\mathrm{TLS}}$ for hBN (red) and control (blue) resonators with error bars showing the standard error of the fit. Box plots show the median, quartiles, and outliers beyond 1.5$\times$ the inter-quartile range. }
    \label{fig:temperature_sweep}
\end{figure*}

\subsection{Temperature Sweep}
Similarly to microwave drive photons, thermal phonons saturate TLS, resulting in decreasing TLS loss with increasing temperature. The loss  at low power is equivalent to the loss determined by the temperature-dependent frequency shift at high power (see Ref.~\cite{Pappas2011,Gao2008} for more details). Resonator spectroscopy scans were taken with the QICK board during a temperature sweep as a method to determine loss complementary to the previously discussed power sweep.

 The fractional frequency shift as a function of temperature ($\delta f(T)/f_0$) is modeled as a sum of contributions from TLS loss $(\delta f(T)/f_0)_{\rm TLS}$ and thermal QP loss $(\delta f(T)/f_0)_{\rm QP}$ ~\cite{Crowley2023}. The TLS loss component is
\begin{equation}
    \label{temp_eq}
\begin{split}
\left( \frac{\delta f(T)}{f_0}\right)&_{\rm TLS} = \frac{F \delta^{0}_{\rm TLS}}{\pi} \times \\
&\left[{\rm Re}\left(\Psi\left(\frac{1}{2} + \frac{1}{2\pi i} \frac{\hbar \omega}{kT}\right)\right) - \log\left(\frac{\hbar \omega}{2\pi kT}\right)\right]
\end{split}
\end{equation}
where $F$ is the filling factor, $\delta^{0}_{\rm TLS}$ is the zero-temperature TLS loss, and $\Psi$ is the complex digamma function. The quantity $F\delta_{\mathrm{TLS}}^{0}$ is an effective loss scaled by the fraction of the electric field that interacts with the dielectric of interest, and therefore involves geometric factors of the thin film thickness and location within the device. The thermal quasiparticle loss component based on Mattis-Bardeen theory is described in other works (see e.g. Refs.~\cite{Crowley2023, Gao2008}), and has free parameters of critical temperature $T_c$ and the kinetic inductance fraction $\alpha$. For superconductors with sufficiently high $T_c$, including Nb, thermal QP loss is negligible at $T_{\mathrm{bath}}  \sim \hslash\omega/(2k_{B})$, $\sim$130\,mK for our resonators, and TLS loss dominates $\delta f(T)/f_0$.  

Figure~\ref{fig:temperature_sweep}a shows ($\delta f/f_0$) of the hBN and control resonators during a temperature sweep spanning the base temperature of the fridge (15\,mK) to 1.1\,K in varying increments. The zero-temperature resonant frequency $f_0$ is estimated as $f_R$ measured at the base temperature. In the range 15--300\,mK, the characteristic hook shape known to arise from TLS and described by Equation~\eqref{temp_eq}  ~\cite{Pappas2011,Gao2008} is observed (Figure~\ref{fig:temperature_sweep}a inset). Near 1\,K, $\delta f(T)/f_0$ trends downwards in both the hBN and the control resonators, resembling emerging dominance of thermal QP loss established by Mattis-Bardeen theory~\cite{Mattis1958}. However, the temperature at which the onset of this negative shift in $\delta f/f_0$ occurs is not consistent with a $T_c$ near 9\,K, as expected for Nb, and fitting the Mattis-Bardeen equations gives a $T_c$ near 2\,K. We note that compressively strained Nb thin films exhibit a reduced $T_c$~\cite{Choi2020}, however, the film stress of the sputtered Nb was not investigated in this work. Therefore, we opt to use only the low end of the temperature sweep for fitting to Equation~\eqref{temp_eq}. Assuming reasonable values of $T_c > 8.5$\,K and $\alpha < 0.1$, the thermal QP contribution is negligible over the entirety of the measured temperature span ($|(\delta f(T)/f_0)_{\rm QP}| < 1.7 \times 10^{-7}$). In further support of setting $T_c > 8.5$\,K, a $T_c$ of 8.9\,K for a different Nb thin film deposited using the same AJA sputtering system and following the same outgassing and seasoning procedure described in Section~\ref{s:methods} was measured using a Physical Property Measurement System. 

\begin{figure*}[tp!]
        \includegraphics[width=0.99\textwidth]{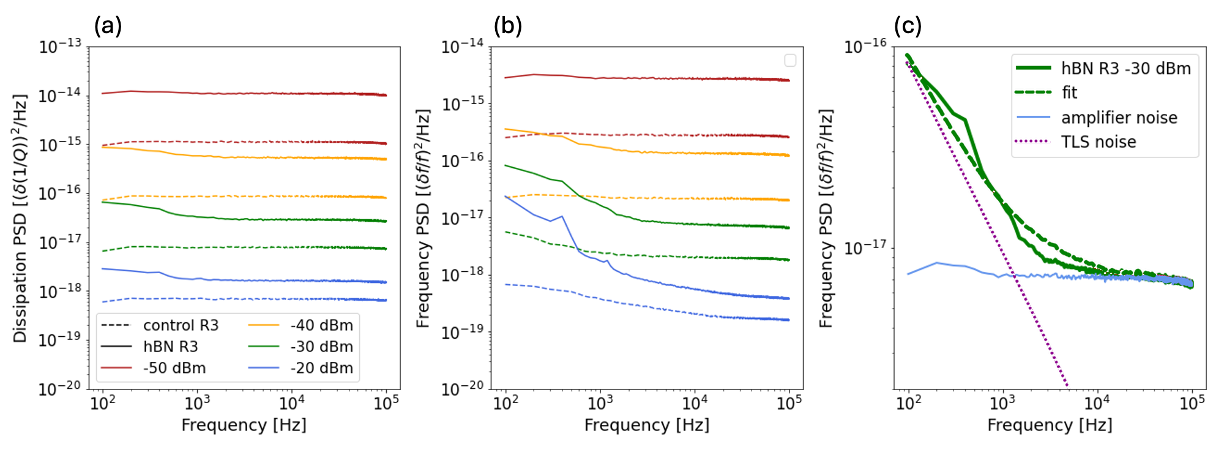}
    \caption{On-resonance noise PSDs of a select hBN resonator (solid lines) and a select control resonator (dashed lines) in the (a) dissipation and (b) frequency quadratures. The line color designates the resonator drive power. (c) Representative fit of frequency PSD to Equation \eqref{noise}, a noise model composed of TLS noise and white amplifier noise components.}
    \label{fig:on_resonance_noise}
\end{figure*}

Figure \ref{fig:temperature_sweep}b shows the $\delta f(T)/f_0$ for select hBN and control resonators (\textit{i.e.}, one of each) up to 300\,mK along with the fit to Equation~\eqref{temp_eq} from which $F\delta^{0}_{\rm TLS}$ is found. The extracted $F\delta^{0}_{\rm TLS}$ of all resonators measured in the temperature sweep using this method are shown in Figure~\ref{fig:temperature_sweep}c. The median $F\delta^{0}_{\rm TLS}$ for the hBN and control resonators are $5.54 \times 10^{-6}$ with standard deviation of $3.9 \times 10^{-7}$ and $3.43 \times 10^{-6}$ with standard deviation of $6.1 \times 10^{-7}$, respectively. Similarly to the measurements of $\delta_i^{\mathrm{LP}}$, $F\delta^{0}_{\rm TLS}$ of the control is close to that of state-of-the-art Nb CPW resonators reported in the literature, for example, $F\delta^{0}_{\rm TLS}$ of Nb-on-Si resonators also extracted from ($\delta f(T)/f_0$) falls within the range $1.8 \times 10^{-6}$ to $2.5 \times 10^{-6}$~\cite{deOry2025}. The presence of hBN increases $F\delta^{0}_{\rm TLS}$ by a factor of $\sim 1.6$.

Power and temperature sweep data yield a consistent picture of dielectric loss in the hBN and control resonators. The control resonators of this study are on par with state-of-the-art Nb CPW resonators, establishing the control/hBN chips as an ideal platform to quantify effects from the presence of hBN. The integration of deposited dielectrics with superconducting circuits generally increases these measures of loss, $\delta_i$ or $F\delta^0_{\rm TLS}$, to order $10^{-5}$ to $10^{-2}$. 
For example, microstrip resonators integrating amorphous hydrogenated SiC have a dielectric loss tangent of $3.1 \times 10^{-5}$~\cite{Buijtendorp2022} and LC resonators integrating amorphous hydrogenated SiN have a loss tangent of $2.5 \times 10^{-5}$~\cite{Paik2010}. Similarly, lumped element resonators using amorphous hydrogenated Si as the dielectric in a parallel plate capacitor have demonstrated TLS noise levels subdominant to that of fundamental noise sources~\cite{Sunil}.
We conclude the presence of MBE-grown, epitaxial, few-layer hBN introduces minimal dielectric loss, with Nb on hBN resonators maintaining $\delta_i^{LP}$ or $F\delta^0_{\rm TLS}$ on the order $10^{-6}$.

\subsection{Finite Element Method Simulations}
Measures of effective loss are practical for quantifying the impact of the hBN thin films on the device performance, but do not directly reveal the dielectric loss of the hBN itself. The material volume and location within the device are baked in. For example, some studies report $F\delta^0_{\rm TLS}$ for an order 10--100\,nm thickness layer of dielectric deposited on top of a CPW resonator~\cite{Wisbey2019}. While the few-layer hBN thin films occupy a smaller volume, the participation ratio of the metal-substrate interface is generally 100$\times$ higher than the metal-air interface~\cite{Martinis2022}.

The total loss can be expressed as a sum over $j$ regions of the loss ($\delta_j$) in each region (e.g. substrate, deposited dielectric, and metal-air (MA), metal-substrate (MS), and substrate-air (SA) interfaces) multiplied by the participation ratio ($p_j$) of each region. Previous works back out $\delta_j$ of a region of interest using $p_j$ from finite element modeling of the device geometry~\cite{Kopas2022,Wang2015}. However, deconvolving the loss contribution from surface oxides and the deposited dielectric is a known challenge for obtaining accurate $\delta^{0}_{\rm TLS}$, independent of $F$, from CPW resonators. Techniques which leverage isolation of the E-field within parallel plate capacitors (PPCs) are better suited to report $\delta_{TLS}^0$, but would require a comparable-quality epitaxial growth of hBN on a superconductor. 
We obtained $p_j$ from finite element method solutions of the eigenmodes of Maxwell's wave equation using AWS Palace and estimated $\delta^{0}_{TLS}$ arising from the hBN interfaces and thin film. Parameters were solved on the geometry of a select resonator (R5) and using bulk dielectric constants for EFG sapphire reported in Ref.~\cite{Read2023}. The surface dielectric constants were set to 10.0 for each of the SA, MS, and MA interfaces as in Ref.~\cite{Deng2022}. The thickness of the MA interface was set to 3 nm, representing the lossy native oxide that forms on Nb. The thickness of the MS interface of the control device was set to 3 nm. We approximate the presence of the hBN as a 6 nm MS interface, lumping together the hBN-sapphire interface, hBN-Nb interface, and the hBN thin film itself. The calculated $p_j$ for the MA, SA, and MS regions of the control resonator were $3.1 \times 10^{-6}$, $1.0 \times 10^{-3}$, and $0.34 \times 10^{-3}$, respectively. Using the median measured $F\delta_{0}^{TLS}$ of the control resonators and assuming a $\delta_{SA}$ of $1.5 \times 10^{-3}$ for EFG sapphire ~\cite{Read2023} and a $\delta_{MA}$ of $1.0 \times 10^{-3}$ for a Nb oxide, we extract an estimated $\delta_{MS}$ of $5.67 \times 10^{-3}$. The calculated $p_j$ for the MA, SA, and MS regions of the hBN resonator were $3.2 \times 10^{-6}$, $1.0 \times 10^{-3}$, and $0.71 \times 10^{-3}$, respectively. The estimated $\delta_{MS}$ of the hBN interfaces and thin film was $5.69 \times 10^{-3}$. 
We conducted this exercise to better inform the comparison of loss across various device and integrated dielectric geometries, but emphasize the caveats of accurately extracting an isolated materials loss from CPW resonator measurements.

\subsection{Time Dynamics}
Dielectric loss gives insight to the time-averaged effect of TLS in the environment of the resonator. The random, temporal fluctuations of the TLS ensemble additionally contribute to noise and instability of the resonator parameters. TLS with energy splittings less than the thermal energy of the environment ($< k_{B}T$), designated as two-level fluctuators, incoherently flip between energy levels on experimental timescales and contribute to the low-frequency noise spectrum~\cite{Mller2019}. TLS with energy splittings $> k_BT$ occupy one of (or a superposition of) their eigenstates on experimental timescales. These TLS are strongly coupled and may resonantly absorb energy from the resonator, which modifies the dielectric constant and induces shifts in $f_R$ and $Q_i$~\cite{Mller2019}. We investigated timestreams of the resonators on fast (1\,Msps) and slow (0.2\,sps) timescales to capture temporal effects of TLS.

\begin{figure*}[tp!]
        \includegraphics[width=0.99\textwidth]{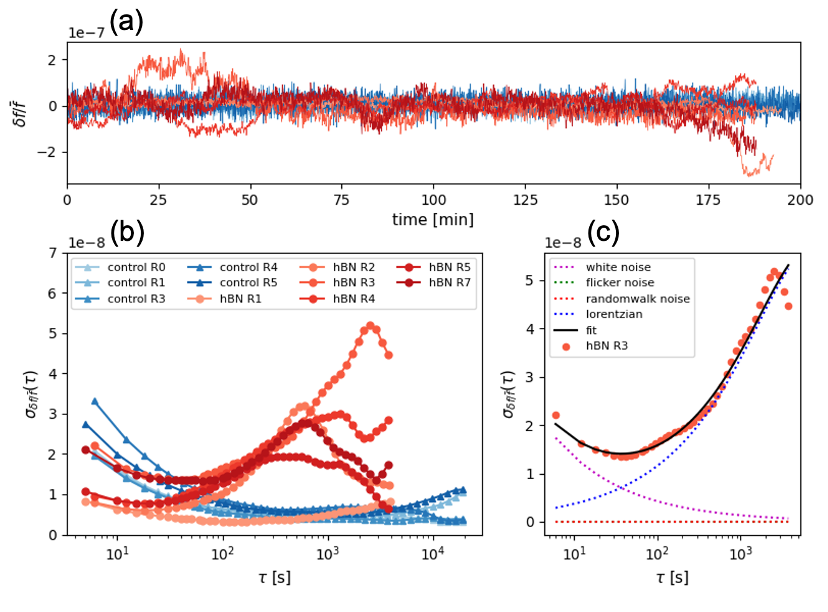}
    \caption{ Fractional frequency shift $\delta f/\bar{f}$ of hBN (red) and control (blue) resonators. (a) 200 minutes of timestream of $\delta f/\bar{f}$ where $f_R$ is extracted from the DCM fit of a resonator spectroscopy scan (b) Allan deviation calculated from the $\delta f/\bar{f}$ timestreams where $\tau$ is the length of the sample window (c) Fit to model with white noise, $1/f$ noise, and random-walk noise components and a Lorentzian term representing the influence of TLS. }
    \label{fig:long_timestream}
\end{figure*}

Figure \ref{fig:on_resonance_noise} shows the power spectral density (PSD) of the on-resonance noise of one of each of the hBN and control resonators collected using the USRP streaming DAQ at drive powers ranging from -50 dBm to -20 dBm at the fridge input. First, $f_R$ was identified from a resonator spectroscopy scan. Next, the $S_{21}$ parameter was recorded at a sample rate of 100\,Msps and decimated by a factor of 100 for the duration of a 60\,second on-resonance probe tone. The first 100\,$\mu$s of the timestream was cut to remove the initial resonator ring-up time. The $S_{21}$ data was transformed to the frequency and dissipation quadratures,~\cite{Temples2024} and the Welch estimate of the PSD was computed. 

The dissipation PSDs (Figure \ref{fig:on_resonance_noise}a) of both the hBN and control resonators were mostly flat, indicative of white amplifier noise. However, we note a slight slope on the low-frequency end of the hBN dissipation PSD. In contrast, the frequency PSDs (Figure \ref{fig:on_resonance_noise}b) demonstrated a more significant low-frequency slope emerging above the white noise floor. Considering electronics noise is correlated in the dissipation and frequency quadratures, this evidences an additional noise source beyond electronics or white noise. The departure from white noise in the low-frequency regime of the frequency PSD is further demonstrated by the power scaling of the PSDs sampled at select frequencies (see Figure \ref{fig:power_scaling_at_freq} of Appendix \ref{sec:app4}). At 25 kHz, the noise of the control and hBN resonators was flat and followed  $P^{-1.1}$ and $P^{-1.3}$ scaling, respectively, in both quadratures, with $P^{-1}$ scaling expected for white noise. At 0.2 and 1 kHz, the power scaling diverges towards a lower power law exponent in the frequency quadrature. 
 
TLS noise predominantly presents in the frequency quadrature; we proceeded with decomposition of the frequency PSD $J(f)$ into white and TLS noise $J_{\mathrm{TLS}}(f)$ components as described by the following model:~\cite{Temples2024}   
\begin{equation}
\label{noise}
J(f) = A_{\rm TLS} f^{n} \left|\frac{1}{1+2jQ(f/f_R)}\right|^2 + A_{w}J_{w}(f),
\end{equation}
where $A_{\rm TLS}$ is the magnitude of the TLS noise power, $n$ is the TLS frequency-dependence exponent, $A_w$ is the magnitude of the white amplifier noise power which is fixed by the power scaling extracted at 25 kHz, and $J_{w}(f)$ is taken as the lowest-power frequency PSD (-50 dBm). The TLS noise component is modified by the resonator response function which rolls off at a frequency of $f_R/2Q$. 
A representative fit of the frequency PSD to Equation \eqref{noise} is shown in Figure \ref{fig:on_resonance_noise}c demarcating the white and TLS noise components.

Figure \ref{fig:power_scaling_TLS} of Appendix \ref{sec:app5} shows the magnitude of the TLS noise component sampled at 1 kHz for the powers at which TLS noise emerged above the white noise background and therefore yielded an appropriate fitting to Equation \eqref{noise}. The TLS noise component for the control and hBN resonators demonstrated scaling of $P^{-0.51}$ and $P^{-0.59}$, respectively. The expected power scaling for TLS noise is $P^{-0.5}$, further supporting a reasonable assumption of TLS noise~\cite{Zmuidzinas2012}. The TLS noise component of the select hBN resonator is about 1.5 orders of magnitude larger than the select control resonator. 
Across the set of measured hBN and control resonators, the mean TLS noise component at 1 kHz and -30 dBm power at the fridge input ($\langle$n$\rangle \sim10^4$) is $4.65 \times 10^{-18}$ and $8.16 \times 10^{-19}$, respectively.   

In order to investigate TLS activity on long timescales, a few hour timestream of the fractional frequency shift $\delta f/\bar{f}$ of the control and hBN resonators was constructed from resonator spectroscopy measurements $~$ 5 seconds in duration taken with the QICK board, where $\bar{f}$ is an arbitrary reference point taken as the mean of $f_R$ over the timestream. The drive power was -20 dBm which saturates the majority of two-level fluctuators. Figure \ref{fig:long_timestream}a shows the $\delta f/\bar{f}$ timestream, demonstrating that most of the hBN resonators exhibit pronounced frequency switching, while the control resonators remain steady.  Figure \ref{fig:long_timestream}b plots the Allan deviation of the $\delta f/\bar{f}$ timestream as a function of the sampling window $\tau$. On the high-frequency (low $\tau$) end, we observe the $\tau ^{-\frac{1}{2}}$ shape characteristic of white noise. 
The larger magnitude of white noise in the control resonators may arise from a warm amplifier that was installed following a power splitter on the RF input line, present for the control but not hBN measurements. 
We note the TLS noise component is extracted from the slope of the data. A model of the Allan deviation is adopted from (Ref.~\cite{Yang2023}), which sums contributions from white noise, $1/f$ noise, and random-walk noise, and assigns Lorentzians to capture the presence of strongly-coupled TLS. As shown in Figure \ref{fig:long_timestream}c, the features in the Allan deviation of the hBN resonators are Lorentzian-like and do not align with the other known noise components. While the use of this model serves as evidence of frequency switching induced by TLS, we note it does not produce a good fit to the Lorentzian-like features themselves across the set of hBN resonators, which perhaps could be better described by a sum of many Lorentzians. We observed a correlation between Lorentzian-like features in the Allan deviation of $\delta f/\bar{f}$ and of the fractional $Q_i$ shift of the hBN resonators, which is consistent with TLS activity. 

The timestreams demonstrate increased TLS-driven fluctuations in the hBN resonators compared to the control resonators. The hBN resonators have a larger TLS noise component contribution by roughly one order of magnitude to the on-resonance noise, and undergo more frequency switching over long, few-hour timescales.

\section{Conclusion}
The results establish epitaxial, few-layer hBN as a low-loss dielectric compatible with superconducting circuits, and supports the feasibility of leveraging the thermal transport properties of hBN for quasiparticle burst suppression without significantly degrading device performance.  While this work utilizes Molecular Beam Epitaxy, the more accessible method of Chemical Vapor Deposition has demonstrated wafer-scale, wrinkle-free, epitaxial few-layer hBN of comparable quality, representing a scalable option for integration of hBN ~\cite{Jang2016}.  Our intended use-case requires hBN at the superconductor-substrate interface. hBN may also be integrated between superconducting leads, for example, exfoliated hBN flakes have been used as tunnel barriers or parallel plate capacitor dielectrics. A follow-up to this work could be the pursuit of epitaxial hBN thin films on a superconducting thin film for a scalable, low-loss dielectric in superconducting junctions. Furthermore, the results inform fabrication of superconducting circuits which integrate 2D materials to leverage their topological functionalities.

\section*{Acknowledgments}
J.C.-S. led this work and was supported by the U.S. Department of Energy, Office of Science, Office of Workforce Development for Teachers and Scientists, Office of Science Graduate Student Research (SCGSR) program. The SCGSR program is administered by the Oak Ridge Institute for Science and Education for the DOE under contract number DE-SC0014664. 
D.J.T. , R.L., and S.S. provided crucial mentorship and guidance throughout all stages of the measurement and analysis. D.J.T. set up the experimental hardware and produced the code for data collection and analysis with the USRP streaming DAQ. A.M. assisted with the device fabrication. S.S. assisted with the device packaging. F.A. provided EM simulations used to interpret these results.  
K.A., D.Bo., G.B., E.C., A.H.C.C., R.K., D.M., A.R., O.S., and S.U. provided key support with installation, data-acquisition scripts, and/or cryostat operation. L.A., Y.W., O.P., and D.So. assisted with hBN thin film growth by Molecular Beam Epitaxy.
Z.T., D.B., and E.F.-F. oversaw this work. The remaining authors assisted with code and hardware development and operation of the QUIET resources. 
This work was performed in part at the Cornell NanoScale Facility, a member of the National Nanotechnology Coordinated Infrastructure (NNCI), which is supported by the National Science Foundation (Grant NNCI-2025233). 
This work was funded in part by the Air Force Office of Scientific Research (award no. FA9550-22-1-0177). 
This work made use of the synthesis and electron microscopy facilities of the Platform for the Accelerated Realization, Analysis, and Discovery of Interface Materials (PARADIM), which are supported by the National Science Foundation under Cooperative Agreement No. DMR-2039380. 
This work made use of the Cornell Center for Materials Research shared instrumentation facility. 
This work is supported in part by the National Science Foundation Graduate Research Fellowship Program under Grant No. DGE-2234667. Any opinions, findings, and conclusions or recommendations expressed in this material are those of the authors and do not necessarily reflect the views of the National Science Foundation.
This work is supported in part by the U.S. Department of Energy, Office of Science, Office of Basic Energy Sciences under Award No. DE-SC0026556.
This work is supported in part by the U.S. Department of Energy, Office of Science, Office of High Energy Physics, under the Quantum Information Science subprogram. 
This manuscript has been coauthored by FermiForward Discovery Group, LLC under Contract No. 89243024CSC000002 with the U.S. Department of Energy, Office of Science, Office of High Energy Physics.

\bibliographystyle{apsrev4-1}
\bibliography{hbn_resonator}

\clearpage
\appendix

\onecolumngrid
\begin{center}
{\bf \large Appendix}
\vspace{12pt}
\end{center}
\twocolumngrid

\section{Molecular Beam Epitaxy Parameters}
\label{sec:app0}
The hBN thin films were grown by Molecular Beam Epitaxy in a Veeco GEN10 system. (001)-oriented sapphire substrates (Kyocera Company) were chemically cleaned in a warm, sonicated bath of micro-90 set to 65°C, followed by room temperature, sonicated baths of acetone and IPA. An elemental boron source was sublimated by electron-beam evaporation. Prior to the growth, a boron flux of $5.0 \times 10^{13}$~atoms~$\rm cm^{-2}\,s^{-1}$ was measured with a Quartz Crystal Microbalance (QCM), and was maintained during the growth by adjusting the electron-beam power to hold a target electron impact emission spectroscopy (EIES) sensor reading. The N$_{2}$ gas flow rate was set  to maintain a chamber pressure of $9 \times 10^{-6}$\,Torr, and the N$_{2}$ plasma was sparked at 500\,W RF power. A CO$_{2}$ laser with 10.6\,$\mu$m wavelength heated the sapphire substrate to 1300$^\circ$C and the temperature was read out by a pyrometer. The growth duration was three hours. 

\begin{figure}[b!]
        \includegraphics[width=0.49\textwidth]{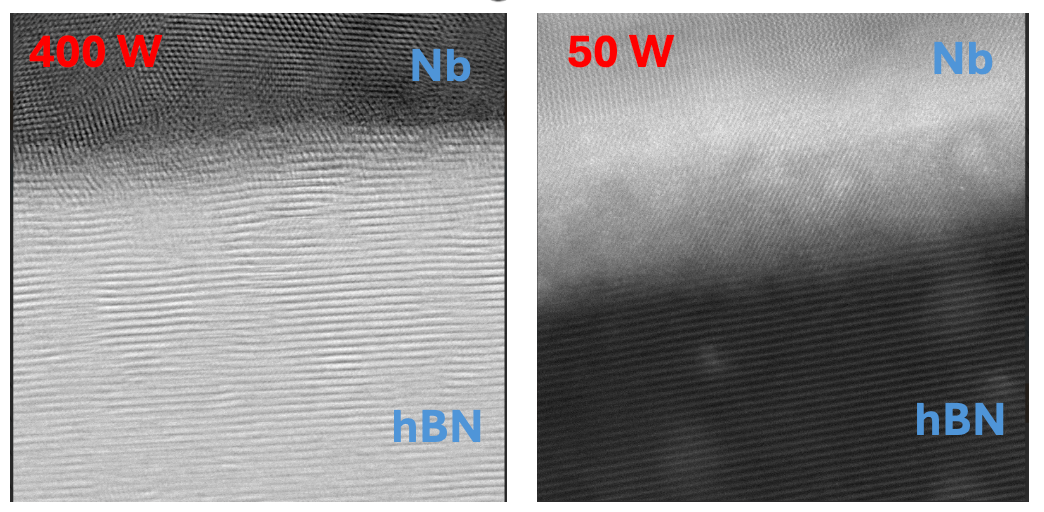}
    \caption{Cross-sectional STEM images of Nb sputtered on exfoliated hBN flakes.  Left: High-power sputtering (400 W) damages the hBN layers extending several nm below the interface. Right: Low-power sputtering (50 W) maintains the ordered hBN layers. Blurriness near interface is an artifact of the FIB cross-sectioning. }
    \label{fig:TEM}
\end{figure}

\section{Resonator Fabrication Parameters}
\label{sec:app1}

\textbf{Sputtering.}
The empty sample carrier and clips were outgassed in the load lock of the AJA sputtering system to $5 \times 10^{-8}$\,Torr. Next, the sample carrier was transferred to the main chamber and was seasoned with a 300\,second deposition of Ti at DC power of 400\,W over a 6\,in diameter target and Ar gas flowrate of 3\,sccm, followed by a 300\,second deposition of Nb using the same parameters. The carrier was removed from the system, loaded with substrates, and returned to the load lock for a second outgassing to $5 \times 10^{-8}$\,Torr. Nb thin films were sputtered with Ar gas flowrate set to 2\,sccm, and the sputtering was performed at room temperature without substrate heating.
Sputtering damage was checked on an exfoliated hBN flake transferred to a sapphire substrate. We opted for hBN flakes because their excellent crystallinity lends itself to easily identifying sputtering damage.Cross-sectional STEM images of the Nb-hBN interface show reducing the sputtering power to 50 W is effective in mitigating sputtering damage to the hBN (Fig. \ref{fig:TEM}).

\begin{figure}[b!]
        \includegraphics[width=0.49\textwidth]{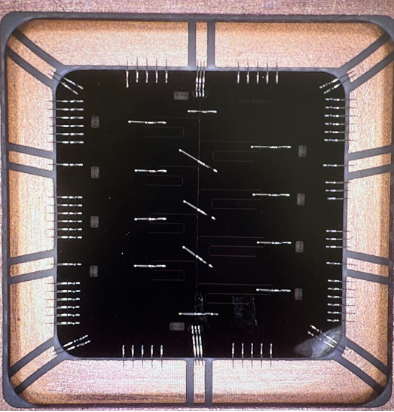}
    \caption{Image of resonator chip wire bonding and packaging.}
    \label{fig:device}
\end{figure}

\textbf{Lithography and Etch.}
Microposit S1805 positive photoresist was spun at 4000\,rpm for 60\,seconds, followed by a 45\,second soft bake at 115$^\circ$C. A Heidelberg DWL2000 mask writer was used to make a chrome mask of the device layout. The photoresist was exposed for 2.6\,seconds to a near-UV light source in an ABM contact aligner tool. The photoresist was developed for 60\,seconds in AZ 726 MIF developer and rinsed with DI water. The Nb was etched using chlorine-based chemistry in a PlasmaTherm PT770 etcher, which employs a combination of reactive ion etching (RIE) and inductively coupled plasma (ICP) etching. The pressure was 7\,mTorr and the gas flow rates were 30:20:5\,sccm Cl$_2$:BCl$_3$:Ar. The RIE power was 12\,W, and the ICP power was 800\,W. After etching, the photoresist was removed in a sonicated bath of Microposit 1165 resist stripper. The clearance of the etch was confirmed by measuring a large resistance (overload) between DC probes positioned on the bond pad and within the trench adjacent to the bond pad where the bare sapphire substrate was exposed. The room temperature resistances from bond pad to bond pad of the hBN and control devices were 5.6\,k$\Omega$ and 5.7\,k$\Omega$, respectively.

\begin{figure}[t!]
        \includegraphics[width=0.49\textwidth]{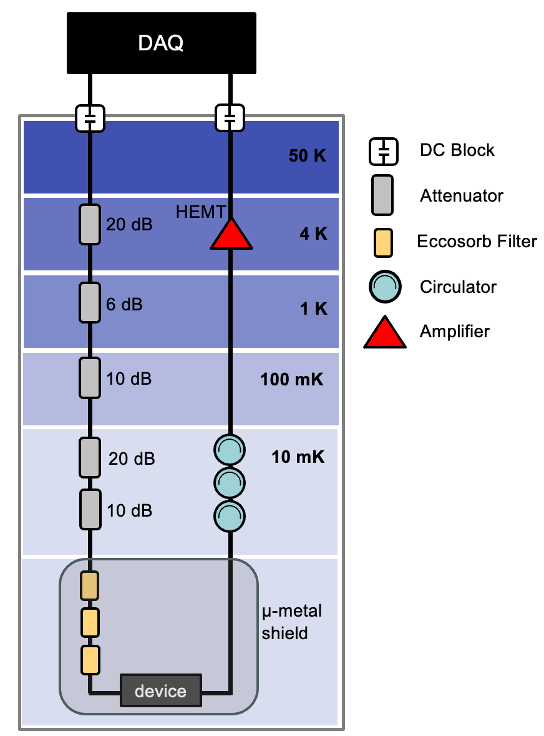}
    \caption{Diagram of fridge and RF input and output lines for resonator measurements.}
    \label{fig:fridge_diagram}
\end{figure}

\section{Packaging}
\label{sec:app2}
SMA connectors were soldered to the device PCB with solder completely filling the gap between the connector and PCB to block stray IR light. The devices were mounted without clamps at the four corners and suspended over a square recess designed to eliminate box modes. A small amount of In wire was placed under each corner to absorb the force imparted by wire bonding. Al wire bonds were formed with an industrial Hesse wire bonder to connect the active PCB trace to the bond pads, the Nb ground plane at the perimeter of the chip to the PCB ground, and to form isolation bonds over the resonator traces and feedline (Fig. \ref{fig:device}).

\section{Fridge Wiring Diagram}
\label{sec:app3}
The devices were installed in an Oxford Proteox-MX dry Dilution Refrigerator. The components on the input and output RF lines at each temperature stage of the fridge are detailed in Figure \ref{fig:fridge_diagram}. The RF lines had stainless steel conductors, with the exception of NbTi lines connecting the RF output from the 10\,mK to 50\,mK stages,  The RF input lines were thermalized at each temperature stage with bulkhead attenuators. The input lines featured three eccosorb foam filters just prior to the device. HEMTs (Low Noise Factory LNF-LNC4\_8G 0920H and LNF-LNC4\_8C 3446H) were installed at the 4\,K stage on the output lines.

\section{Resonator Parameters}
\label{sec:app4}

Table \ref{tab:placeholder} lists the resonant frequencies and coupling quality factors of the hBN and control resonators. These values were measured at base temperature and were consistent across the power sweep.

\begin{table}[b!]
    \centering
\caption{Resonant frequencies and coupling quality factors of hBN and control resonators measured at base temperature.}
\label{tab:placeholder}
    \begin{tabular}{r|cccc}
         Resonator&  $f_R$ [GHz]&  $Q_c$ [$10^5$]&  $f_R$ [GHz] & $Q_c$ [$10^5$]\\
         Index&  (control)&  (control) &   (hBN) & (hBN) \\ 
         \hline
         R0&4.2650  &2.68  &4.2614  &3.39 \\
         R1&4.6035  &2.38  &4.6298  &3.44 \\
         R2&4.9139  &2.12  &5.0119  &3.74 \\
         R3&5.1054  &2.30  &5.3790  &2.24 \\
         R4&5.5094  &1.38  &5.7297  &1.87 \\
         R5&5.9862  &2.04  &6.1051  &2.52 \\
         R6&6.4333  &1.84  &6.4790  &1.77 \\
         R7&6.8352  &2.06  &6.8302  &1.62 \\
    \end{tabular}

\end{table}

\section{On-Resonance Noise Analysis}
\label{sec:app5}

\begin{figure*}[t]
        \includegraphics[width=0.8\textwidth]{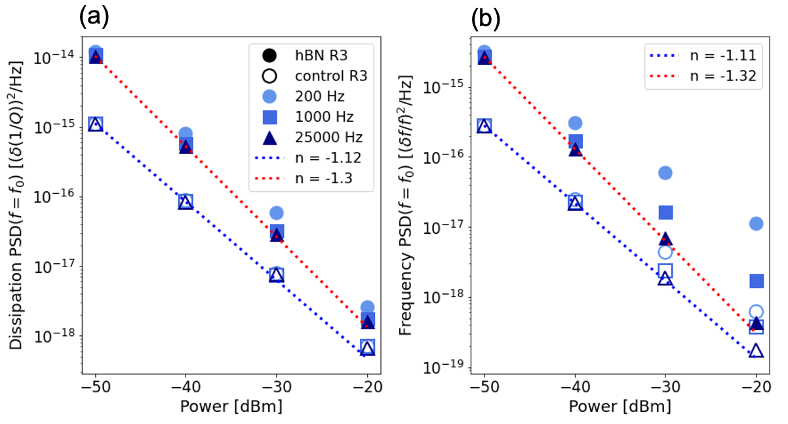}
    \caption{Power-dependence of (a) dissipation and (b) frequency PSDs sampled at 0.2 (circle markers), 1 (square markers), and 25 kHz (triangle markers) for the select hBN (filled markers) and control (open markers) resonators. The dashed lines show the power law fit for the data sampled at 25 kHz of the hBN (red) and control (blue) resonators.}
    \label{fig:power_scaling_at_freq}
\end{figure*}

The power scaling of the white noise magnitude in Equation \eqref{noise} was fixed by fitting a power law to points sampled from the frequency PSDs at 25 kHz (Figure \ref{fig:power_scaling_at_freq}).  

\begin{figure}[t]
        \includegraphics[width=0.4\textwidth]{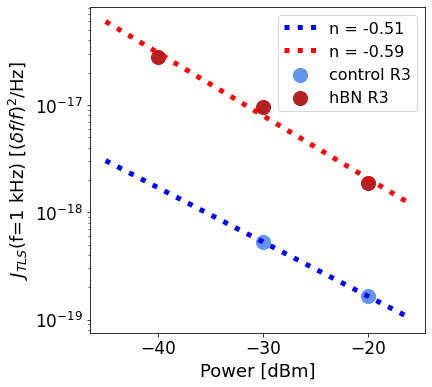}
    \caption{Power-dependence of the magnitude of the TLS noise component sampled at 1 kHz for the control (blue markers) and hBN (red markers) resonators. The dashed lines show the power law fit.}
    \label{fig:power_scaling_TLS}
\end{figure}

The power scaling of the TLS noise extracted from fitting of Equation \eqref{noise} was found by sampling the magnitude of the TLS noise component $J_{\mathrm{TLS}}(f)$ at 1 kHz and fitting a power law (Figure ~\ref{fig:power_scaling_TLS}).

\end{document}